\documentclass[12pt]{article}
\usepackage[utf8]{inputenc}
\usepackage{sgame}

\usepackage{rotating}
\usepackage{float}
\usepackage{multirow}
\usepackage{lscape}
\usepackage{quotes}
\usepackage{cite}
\usepackage{amsfonts,amsmath, amsthm, amssymb}
\usepackage{bbm}
\usepackage{hyperref}
\usepackage{pdfpages}
\usepackage[round]{natbib}
\usepackage{booktabs}
\usepackage{caption,subcaption}
\usepackage[normalem]{ulem}

\usepackage{graphicx}
\usepackage{enumitem}
\usepackage{rotating}
\usepackage{setspace}
\usepackage{soul}

\def\stackbelow#1#2{\underset{\displaystyle\overset{\displaystyle\shortparallel}{#2}}{#1}}

\newtheorem{prop}{Proposition}

\usepackage{geometry}
\usepackage{xpatch}
\makeatletter
\AtBeginDocument{\xpatchcmd{\@thm}{\thm@headpunct{.}}{\thm@headpunct{}}{}{}}
\makeatother

\usepackage{pifont}
\newcommand{\crossmark}{\text{\ding{55}}}%

\newcommand{\R}{\mathbb{R}}

\begin{document}

\title{Credibility in Second-Price Auctions: \\ An Experimental Test}
\author{Ahrash Dianat\thanks{Department of Economics, University of Essex, a.dianat@essex.ac.uk} \and Mikhail Freer\thanks{Department of Economics, University of Essex, m.freer@essex.ac.uk}}
\maketitle

\begin{abstract}

We provide the first direct test of how the credibility of an auction format affects bidding behavior and final outcomes.  To do so, we conduct a series of laboratory experiments where the role of the seller is played by a human subject who receives the revenue from the auction and who (depending on the treatment) has agency to determine the outcome of the auction.  Contrary to theoretical predictions, we find that the non-credible second-price auction fails to converge to the first-price auction.  We provide a behavioral explanation for our results based on sellers' aversion to rule-breaking, which is confirmed by an additional experiment.

%We conduct a laboratory experiment to investigate behavior and outcomes in a non-credible version of the second-price auction.  In contrast to previous experimental studies, the role of the seller is played by a subject who has agency to determine the outcome of the auction and who receives the revenue from the auction.  We find that a large majority of bids lie between the theoretical benchmarks of the first-price auction and the credible second-price auction.  While sellers often break the rules of the auction and overcharge the winning bidder, they typically do not maximize revenue.  We provide a behavioral explanation for our results based on incorrect beliefs (on the part of bidders) and lie-aversion (on the part of sellers), which is confirmed by revealed preference tests.

\end{abstract}

\bigskip



\clearpage

\newpage
\section{Introduction}

The standard framework in auction theory models bidders as strategic agents but assumes that the seller is non-strategic - i.e., that the seller can commit herself to faithfully implementing the rules of the auction.  Recent theoretical work by \citet{akbarpour} relaxes this assumption.  In particular, they allow the seller to deviate from the auction rules in ways that have \textit{innocent explanations} - i.e., in ways that cannot be detected by bidders given the information they have.  They define an auction to be \textit{credible} if there is no scope for the seller to break the rules without being detected.\footnote{In the presence of cheating opportunities, \citet{akbarpour} also define an auction to be credible if it is still incentive-compatible for the seller to follow the rules.  However, the example where incentive compatibility has empirical bite (i.e., the ascending price auction) is outside the scope of this paper.}

As a concrete example, consider the first-price and second-price sealed-bid auctions for the sale of a single item.  It is easy to see that the first-price auction is credible.  The seller cannot cheat without being detected, since the winner knows her own bid and therefore knows how much she should be charged.  The second-price auction, however, is not credible.  After observing all the bids, the seller can exaggerate the second-highest bid and overcharge the winning bidder up to the amount of her own bid.  Moreover, this type of cheating cannot be detected by the winner as long as she does not observe the second-highest bid.  In the language of \citet{akbarpour}, the seller's behavior is supported by the innocent explanation that the highest and second-highest bids are identical.

%Even though there is anecdotal evidence and common sense pointing towards the existence of the credibility issues in real life, there are still couple concerns to be generated. First, the agents do not like to deviate from the rules, even when these rules are non-binding (citations).  Therefore, one should not expect for sure to observe the perfect result as in \cite{akbarpour}, i.e. the non-credible version of second-price auction being equivalent to the first-price auction. Hence, whether sellers have extreme (compliant or shameless) or heterogeneous preferences would generate the level of uncertainty about the rules of the game faced by the bidder. The second problem is to which extent bidders can cope with the appeared complexity, both in terms of best-responding to seller's and other bidders' behavior. Hence, in order to understand the possibility of application of the non-credible mechanisms in practice we need to (1) evaluate the extent to which non-credibility is an issue, (2) evaluate to which extent non-credibility forces bidders to deviate from the equilibrium behavior.

Credibility is not just a theoretical concern, but also a tangible property of auctions that can have significant real-world implications.  Indeed, \citet{akbarpour} provide anecdotal evidence of a seller taking advantage of the lack of credibility in a second-price auction:

\begin{quote}
"An auctioneer running second-price auctions in Connecticut admitted, ‘After some time in the business, I ran an auction with some high mail bids from an elderly gentleman who’d been a good customer of ours and obviously trusted us. My wife Melissa, who ran the business with me, stormed into my office the day after the sale, upset that I’d used his full bid on every lot, even when it was considerably higher than the second-highest bid.’ \citep{lucking}.”
\end{quote}

\noindent The suspicion that second-price auctions are often rigged in this fashion can affect both entry and bidding decisions.  First, bidders may simply choose not to participate in second-price auctions. This would undermine the market design objective of creating sufficiently "thick" marketplaces, and also decrease the expected revenue from running these types of auctions.  Second, bidders who still choose to participate may adjust their bidding strategies accordingly.  This latter point undermines the common argument for the use of second-price auctions: that strategy-proof mechanisms provide a level playing field and save participants from the cognitive/monetary costs of plotting their strategies.  More generally, we would expect the problems arising from non-credible auctions to be exacerbated in settings where bidders receive limited feedback (e.g., sealed-bid auctions), where bidders are not physically present when the auction is conducted (e.g., auctions that allow for online, mail, or telephone bidding), or where reputational concerns play a weaker role (e.g., markets with a single seller or a large number of homogeneous sellers).

However, there are no existing empirical studies on the performance of non-credible auctions.  This is an important gap to fill.  Non-credible auctions may present a challenge for bidders, who must cope with an additional dimension of strategic uncertainty: identifying optimal bidding strategies now requires considering the actions of the seller as well as other bidders.  To that end, it also not obvious to what extent sellers will take advantage of the lack of credibility in auction settings.

In this paper, we bring empirical evidence to bear on these questions by providing the first direct test of how the credibility of an auction format affects bidding behavior and final outcomes.  To do so, we conduct a laboratory experiment in which each subject participates in one of the following three treatments: the first-price auction (FP), the credible second-price auction (CSP), and the non-credible second-price auction (NCSP).\footnote{
It is natural to wonder about a possible fourth treatment: a non-credible version of the first-price auction.  However, this hypothetical treatment would be vacuous given that the first-price auction is inherently self-policing.  The seller cannot deviate from the auction rules without being detected since the winning bidder knows her own bid and therefore knows how much she should be charged.  In the framework of \citet{akbarpour}, there are no "innocent explanations" for seller cheating.  In the real world, we should also not expect the persistence of non-credible first-price auctions since seller cheating (in the form of overcharging) can be detected and punished with certainty.}
In all three treatments, the role of the seller is played by a subject who receives the revenue from the auction.  In the FP and CSP treatments, the seller's role is passive and the outcome of the auction is determined according to the prescribed rules.  In the NCSP treatment, the seller has an active role.  In particular, the NCSP auction is implemented as a dynamic game with two stages: bidders submit their bids in sealed envelopes (after observing their private values), and then the seller determines the outcome of the auction (after observing the bids).

To resolve equilibrium selection, we focus attention on the symmetric Perfect Bayesian equilibrium in undominated strategies.  This provides a sharp theoretical prediction for both sides of the market.  Sellers will select the highest bidder as the winner and choose a price equal to the highest bid.  Bidders will internalize this and bid "as if" they are participating in a first-price auction.  Taken together, the core theoretical prediction is the strategic equivalence of the NCSP and FP auctions.

While our results are qualitatively consistent with the theory, the main theoretical prediction is not borne out. Although sellers in the NCSP auction often break the rules of the auction and overcharge the winning bidder, they typically do not maximize revenue.  Consequently, most bidders in the NCSP auction do not behave as though they are participating in a FP auction.  In fact, the estimated bidding function in the NCSP auction is significantly different than the estimated bidding functions in \textit{both} the FP and CSP auctions.  This suggests that bidders perceive the three auctions as distinct mechanisms from a behavioral perspective.

As a next step, we consider a behavioral model in which sellers are averse to rule-breaking.  We show that the model generates predictions that are consistent with our experimental results.  To test the underlying mechanism of the model, we conduct an additional experimental treatment using a "no-rules" version of the NCSP auction.  The new treatment provides support for aversion to rule-breaking as the driving force behind our results.  In particular, we find that both bidders and sellers react to the absence of auction rules by behaving closer to the theoretical predictions for the FP auction.

%\red{I think that the next paragraph is redundant. We can cut it.}
%Our contribution in this paper is threefold.  First, we theoretically characterize equilibrium behavior in the NCSP auction in the presence of behavioral sellers who are averse to rule-breaking.  Our second contribution is methodological: to the best of our knowledge, we are the first to implement a non-credible auction in the laboratory by having a subject play the role of an active seller. Finally, and most importantly, we provide the first empirical study on the performance of non-credible auctions.  Our results suggest that credibility is a necessary condition for realizing many of the practical benefits of second-price auctions.

The paper that is closest to ours in terms of design is \citet{bartling}, who also conduct an experimental investigation of second-price auctions in the presence of human sellers.  However, since the focus of their experiment is on how social preferences affect bidding behavior, sellers only receive the revenue from the auction but do not have agency to determine the outcome of the auction (which is the novel feature of our design).  Our paper also replicates and extends well-known findings from the experimental auction literature.\footnote{For a survey of this literature, see Chapter 7 of \citet{roth}.}  In particular, our results have a natural connection to those reported in the seminal paper of \citet{kagel}.  Consistent with \citet{kagel}, we find that overbidding relative to the risk-neutral Nash equilibrium prediction is common in the FP auction and overbidding relative to the dominant-strategy prediction still occurs in the CSP auction.  Our experiment also demonstrates that bidding behavior varies between two theoretically equivalent institutions (FP and NCSP auctions).  This finding mirrors the more general failure of strategic equivalence between static and dynamic auction procedures that was famously reported in \citet{kagel}.  Finally, our paper joins a significant body of experimental work on the performance of auctions in the presence of behavioral agents - e.g., regret-averse bidders in \citet{ozbay} and level-$k$ bidders in \citet{crawford}.

Although our main contribution is experimental, our paper is also thematically related to the literature on mechanism design with limited commitment.  In particular, \citet{baliga} and \citet{bester} study environments where the designer has agency to determine the outcome of the mechanism, which is a feature that is reflected prominently in our experiment.  In more general allocation problems, \citet{hakimov} introduce the related concepts of "verifiability"  (i.e., mechanisms that allow participants to check if their assignments are correct) and "transparency" (i.e., mechanisms in which the designer cannot cheat without being detected).  Taken together, these concepts are stronger than the notion of credibility we investigate in this paper.  The issue of seller cheating has also drawn significant attention in the field of industrial organization - e.g., the practice of "shill-bidding" by sellers in second-price auctions \citep{mcadams, harstad, porter}.

The rest of the paper proceeds as follows.  Section 2 introduces the theoretical framework, Section 3 provides details of the experimental design, Section 4 presents the experimental results, Section 5 investigates the underlying mechanism driving our experimental results, and Section 6 concludes.

\section{Theoretical Background}
\label{sec:theory}

We present a simple operationalization of the environment in \cite{akbarpour}.  There is a single, indivisible item for sale.  Let $N \cup \lbrace 0 \rbrace$ be a finite set of agents, consisting of $\vert N \vert = n$ bidders and a seller (player 0).  Let $X$ be a set of outcomes, where an outcome $x = (y, t)$ consists of a winner $y \in N$ and a profile of payments $t \in \mathbb{R}^n$ such that $t_i = 0$ for $i \neq y$.  A type space is $\Theta = \times_{i\in N}\Theta_i$, where $\Theta_i = [\underline v, \bar v]$.  Let $F: \Theta_i \longrightarrow [0,1]$ denote a marginal distribution over types.\footnote{ 
    We further assume that $F$ is strictly increasing in its interior and continuously differentiable.
}  
The distribution $F$ is common knowledge.  
The bidding function of player $i$ is given by $b_i: \Theta_i \longrightarrow \R_+$.

After observing the profile of bids $b = (b_1, b_2, \dots, b_n)$, the seller selects the winner $y \in N$ and the price of the item $t_y \in [0, b_y]$.\footnote{ 
    The restriction that $t_y \leq b_y$ captures the assumption in the \citet{akbarpour} model that the seller can only deviate from the rules in ways that have "innocent explanations."
}  
The feedback that bidders receive is captured by a partition $\Omega_i$ of $X$ for each $i \in N$, representing what bidder $i$ directly observes about the outcome.\footnote{
That is, every element of $\Omega_i$ denoted by $\omega_i\subseteq X$ is a subset of the space of outcomes such that $\bigcup_{\omega_i\in \Omega_i} \omega_i = X$ and $\omega_i \cap \omega_i'=\emptyset$ for every $\omega_i,\omega_i'\in \Omega_i$.
}
In particular, $(y, t), (y^\prime, t^\prime) \in \omega_i$ if and only if one of the following two conditions holds: (1) $y \neq i$ and $y^\prime \neq i$, or (2) $y = y^\prime = i$ and $t_i = t_i^\prime$.  
That is, each bidder observes whether she wins the item and observes her own payment.

We now describe the preferences of the players.  Bidders have private values: $u_i (x,\theta) = \mathbbm{1}_{\lbrace{i = y}\rbrace}(\theta_i - t_i)$ for player $i \in N$.  The seller receives the revenue from the auction: $u_0(t_y,b) = t_y$.  We focus attention on the symmetric Perfect Bayesian equilibrium in undominated strategies.  Let $\mathbf{b^{FP}}$, $\mathbf{b^{NCSP}}$, $\mathbf{b^{CSP}}$ denote the symmetric and strictly increasing equilibrium bidding functions in the first-price, non-credible second-price, and credible second-price auctions, respectively.

\begin{prop}
\label{prop:SelfishEquilibrium}
In equilibrium, 
\begin{itemize}
    \item [(i)] the seller selects the highest bidder as the winner.
    
    \item [(ii)] chooses a price equal to the highest bid.  That is,
$$
t^*_y = \max{\lbrace{b_1, b_2, \dots, b_n}\rbrace}.
$$
    
    \item [(iii)] for each bidder $i \in N$ and each value $\theta_i \in \Theta_i$, we have that \\
\begin{equation*}
\mathbb{E} \Big[\max_{j \neq i} \theta_j  \Big\vert \max_{j \neq i} \theta_j < \theta_i \Big] = \mathbf{b^{FP}}(\theta_i) = \mathbf{b^{NCSP}}(\theta_i) < \mathbf{b^{CSP}}(\theta_i) = \theta_i.
\end{equation*}
\end{itemize}
\end{prop}

We omit the proof of Proposition \ref{prop:SelfishEquilibrium} since it follows directly from the results presented in \cite{akbarpour}.  However, we note that Proposition \ref{prop:SelfishEquilibrium} provides a clear testable prediction that the NCSP auction is theoretically equivalent to the FP auction.

\section{Experimental Design}
\label{sec:design}

Our experiment consists of three players (two buyers and one seller) participating in a sealed bid auction for a single fictitious commodity.  Each play of the auction is a round.  Subjects play 10 rounds in an experimental session.  At the beginning of the experiment, each subject is randomly assigned to the role of either a buyer (2/3 chance) or a seller (1/3 chance).  Subjects' roles are fixed across all 10 rounds, but subjects are randomly and anonymously re-matched into groups of three at the start of each round.  We use a hypothetical currency (tokens) to keep track of subjects' earnings throughout the experiment.  Each subject starts the experiment with a token balance: buyers start with 100 tokens and sellers start with 0 tokens.  At the end of a round, each subject's earnings are added to her token balance.  At the end of the experiment, each subject is paid the sum of her final token balance (where 20 tokens = £1) and her payoff from two elicitation tasks.\footnote{We elicit risk attitudes using a task from \citet{potters} and we elicit a measure of lying aversion using a task from \citet{gneezy}.}

We now describe the general structure of the experiment.  At the beginning of a round, each buyer is randomly given a value for the item.  The values are drawn independently and uniformly from the set of integers between 0 and 100 (inclusive).  The buyers' values are private information.  Each buyer independently and simultaneously submits a bid for the item.\footnote{
    The minimum possible bid is 0 tokens and the maximum possible bid is 100 tokens.  Buyers are allowed to bid any amount in this interval, regardless of their values.}

The outcome of the auction depends on the treatment.  We use a between-subject experimental design where each subject participates in one of the following three conditions:

\begin{enumerate}
    \item Credible second-price auction (CSP): The buyer who submits the highest bid wins the item.  The price of the item is equal to the amount of the lowest bid.
    \item Non-credible second-price auction (NCSP): The seller can select either buyer as the winner and can set the price of the item to be any amount that is not greater than the winner's bid.
    \item First-price auction (FP): The buyer who submits the highest bid wins the item.  The price of the item is equal to the amount of the highest bid.
\end{enumerate}

\noindent In all three conditions, each buyer observes whether she won the item.  Only the winner and the seller observe the price of the item.  The winner receives her value for the item, minus the price of the item.  The seller receives the price of the item.

There are a few important points to make about our design choices.  First, since we are interested in how the credibility of an auction format affects bidding behavior, we need to eliminate seller reputation as a possible explanation for our results.  Our random and anonymous re-matching protocol does precisely this: buyers are unaware of sellers' identities and whether they have interacted with a particular seller in a previous round. 
Second, an important feature of our experiment is that a subject always plays the role of the seller and receives the revenue from the auction.  We maintain this feature even in the two auctions with passive sellers (CSP and FP), which allows us to rule out social preferences or fairness concerns as possible explanations for differences in bidding behavior in the NCSP auction.  Third, our experimental auction markets consist of one seller and only two buyers.  This is the simplest possible setting in which credibility plays a role, and thus provides a clean test of the theory.  We do not view this as overly restrictive or harmful to external validity, since many real-world auctions also feature a small number of active bidders.\footnote{\citet{agranov} note that, in 2013, 27\% of eBay auctions had only two bidders while 77\% of eBay auctions had five or fewer bidders.}  Finally, we acknowledge that it is a challenge to successfully induce a non-credible auction in the laboratory.  In particular, it is crucial that both the rules of the auction and the seller's strategy space are common knowledge.  Our experimental instructions emphasize both of these features.\footnote{The instructions for all treatments are provided in the Online Appendix.}  We also include a quiz to verify subjects' comprehension of the instructions.  The experiment does not begin until each subject correctly answers all the quiz questions.

\subsection*{Implementation}

The experimental sessions were run at the Laboratory for Economic and Decision Research (LEDR) at the University of East Anglia.  For the main set of experiments, we recruited a total of 285 subjects across 13 sessions (with either 18, 21, or 24 subjects per session).  Each session lasted approximately 60 minutes.  The experiment was programmed and conducted with the z-Tree software \citep{fischbacher}.  Table~\ref{table:summarystatistics} provides more detailed summary statistics.

\begin{table}[tp]
    \centering
    \resizebox{1\linewidth}{!}{
    \begin{tabular}{ccccc}
    \toprule
        Treatment & Active Seller & \# of Sessions & \# of Subjects & Average Earnings   \\
         \midrule
         First-Price (FP) & $\crossmark$ & 3 & 72 & \textsterling 19.59  \\
         Credible Second-Price (CSP) & $\crossmark$ & 3 & 69 & \textsterling 19.62  \\
         Non-Credible Second-Price (NCSP) & $\checkmark$ & 7 & 144 & \textsterling 19.61 \\
         No-Rules (NR) & $\checkmark$ & 4 & 72 & \textsterling 18.40 \\
         \midrule
         All & & 17 & 357 & \textsterling 19.36 \\
         \bottomrule
    \end{tabular}
    }
    \caption{Overview of experimental design}
    \label{table:summarystatistics}
\end{table}

\section{Experimental Results}
\label{sec:results}

\subsection*{Bidder Behavior}

We begin by investigating bidding behavior.  For the case of two risk-neutral bidders with values independently and uniformly drawn from $[0,100]$, the symmetric Perfect Bayesian Equilibrium (PBE) predictions in undominated strategies are given by

\begin{equation*}
b^{FP}(\theta_i) = b^{NCSP}(\theta_i) = 0.5\theta_i < \theta_i = b^{CSP}(\theta_i).
\end{equation*}

\begin{figure}[tp]
    \centering
    \includegraphics[scale=0.22]{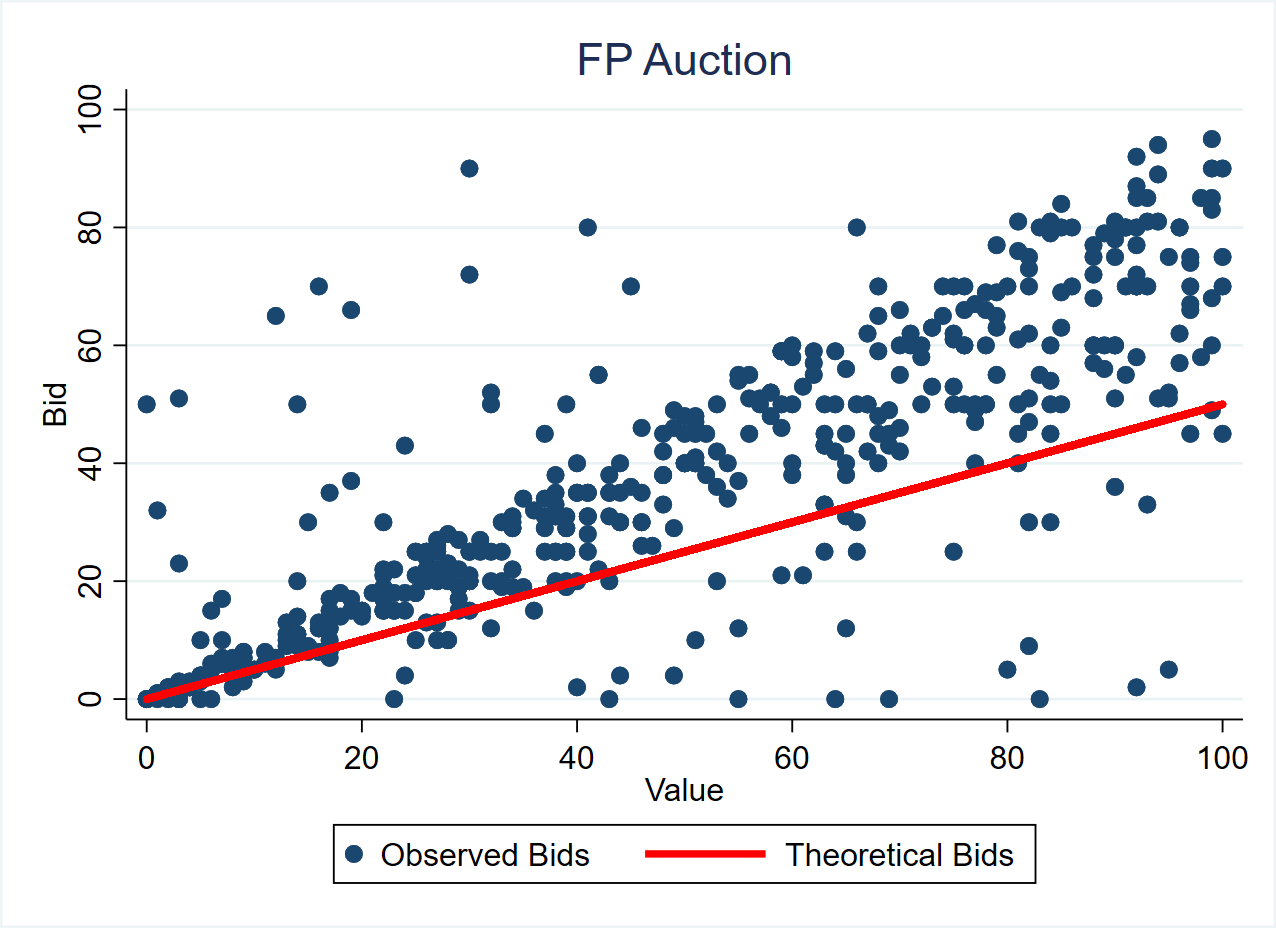}
    \includegraphics[scale=0.22]{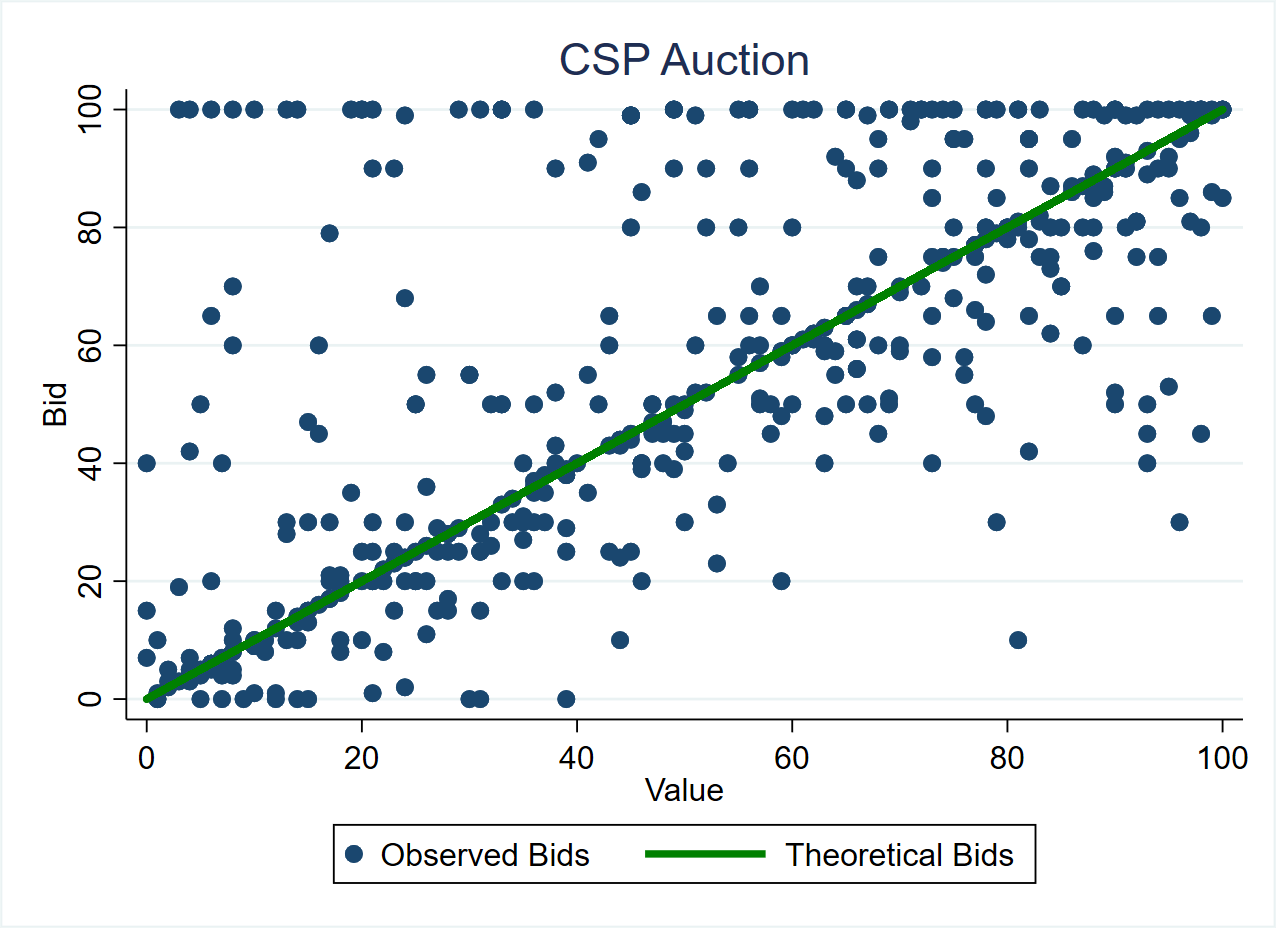}
    \includegraphics[scale=0.22]{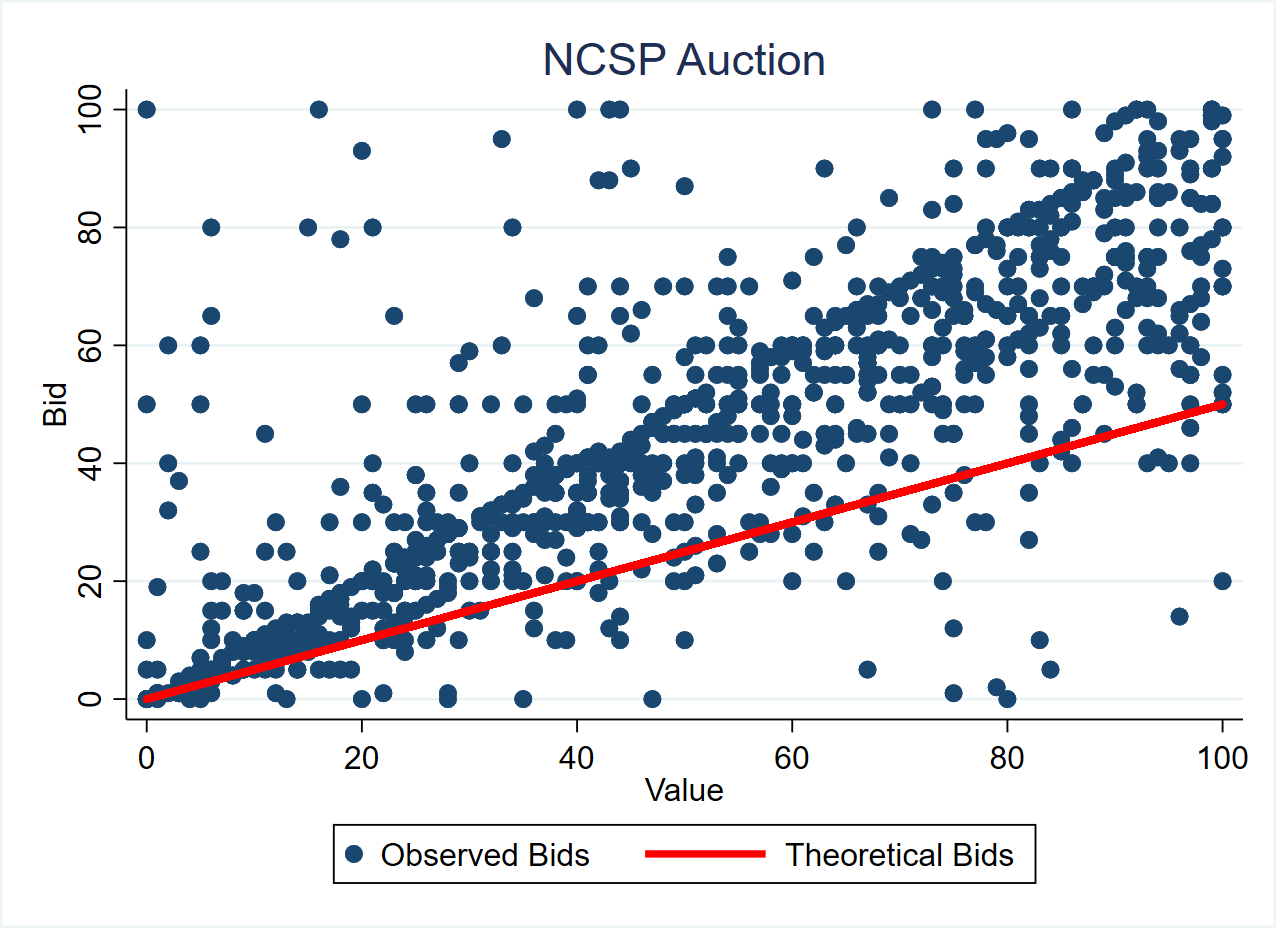}
    \caption{Observed and theoretical bids}
    \label{fig:bids}
\end{figure}

Figure~\ref{fig:bids} shows scatter plots of observed bids/values in the three treatments, along with the theoretical predictions in solid lines.  It is clear that overbidding relative to the equilibrium prediction is common in the FP and NCSP auctions, occurring in 86\% (411/480) and 89\% (853/960) of bid-value pairs, respectively.  In the CSP auction, however, subjects have a dominant strategy of bidding their true values.  Although overbidding still occurs in 39\% (181/460) of cases, 18\% (85/460) of cases correspond to dominant-strategy play.  If we relax this condition to include subjects who bid either 1 token below or 1 token above their true values, then 28\% (127/460) of cases correspond to dominant-strategy play.

As a next step, we conduct linear estimations of the bidding functions (forcing the intercepts to pass through zero).\footnote{We suppress the intercept terms for simplicity and to allow for direct comparisons with the equilibrium bidding functions. The same specification has been used before, for instance in \cite{ozbay}.  The qualitative results remain unchanged if we do not suppress the intercept terms.} Regression 1 in Table~\ref{table:OLS} shows the estimated bidding functions for the three treatments, with the NCSP auction as the baseline.  In particular, we have that

\begin{equation*}
    \stackbelow{\hat{b}^{FP}(\theta_i)}{0.74\theta_i} <^{***} \stackbelow{\hat{b}^{NCSP}(\theta_i)}{0.83\theta_i} <^{***} \stackbelow{\hat{b}^{CSP}(\theta_i)}{1.01\theta_i}.
\end{equation*}

\begin{table}[tp]
\centering
\begin{tabular}{lccccc} \hline
& \multicolumn{2}{c}{All Treatments} & FP & CSP & NCSP\\
& Rounds 1 - 10 & Rounds 6 - 10 & & & \\
Dependent Variable: Bid & (1) & (2) & (3) & (4) & (5) \\ \hline
Value & 0.834*** & 0.845*** & 0.757*** & 0.984*** & 0.845*** \\
& (0.015) & (0.019) & (0.035) & (0.038) & (0.028) \\
Value*$\mathbbm{1}_{\lbrace{FP}\rbrace}$ & -0.097*** & -0.106** & & & \\
& (0.027) & (0.031) & & & \\
Value*$\mathbbm{1}_{\lbrace{CSP}\rbrace}$ & 0.180*** & 0.252*** & & & \\
& (0.029) & (0.036) & & & \\
Value*$\mathbbm{1}_{\lbrace{Risk-Averse}\rbrace}$ & & & -0.025 & 0.045 & -0.015 \\
& & & (0.044) & (0.050) & (0.033) \\
& & & & & \\
Observations & 1900 & 950 & 480 & 460 & 960 \\ 
Number of subjects & 190 & 190 & 48 & 46 & 96 \\ \hline
\end{tabular}
\caption{Linear estimates of the bidding functions (passing through the origin) \\ Standard errors are clustered at the subject level \\ Significance levels are indicated as follows: ** $p < 0.05$, *** $p < 0.01$}
\label{table:OLS}
\end{table}

\noindent Notice that the estimated slopes of the bidding functions are significantly different at the 1\% level.  This generates two main insights.  First, contrary to theoretical predictions, the NCSP auction fails to converge to the FP auction.  Second, the NCSP auction is still behaviorally distinct from the CSP auction.  Overall, bidders' behavior is consistent with the belief that sellers in the NCSP auction are choosing an intermediate price between the highest and second-highest bids.  We will return to this intuition later when we investigate seller behavior.

We conduct several robustness checks.  First, we account for the possibility of subject learning over time.  Regression 2 in Table~\ref{table:OLS} estimates the bidding functions only using data from the last five rounds of the experiment.  Our qualitative results remain unchanged.  Second, we investigate the effect of risk preference on bidding behavior.\footnote{To elicit risk attitudes, we use an investment task from \citet{potters}.  In particular, each subject receives a £2 endowment and decides how much to invest in a risky project that has a 50\% chance of success.  If the project is unsuccessful, the subject loses the amount invested.  If the project is successful, the subject receives three times the amount invested.  Therefore, investing any amount less than £2 is a sufficient condition for risk-aversion.  In our experiment, we can classify 68\% (129/190) of bidders as risk-averse.}  We do this separately for each treatment, since the interplay between risk preference and bidding behavior varies across the different auction formats.  The results are shown in Regressions 3 - 5 in Table~\ref{table:OLS}.  In all three auctions, we find that risk-averse subjects do not bid significantly differently than other subjects.\footnote{
    \textcolor{black}{Although this result might seem surprising, there is a broader experimental literature suggesting that measures of risk preference are not stable across different tasks or contexts \citep[e.g.][]{dohmen,einav,barseghyan,anderson,reynaud}.
    }
}  
This finding is consistent with the theoretical prediction for the CSP auction (where subjects have a dominant strategy), but inconsistent with the theoretical predictions for the FP and NCSP auctions (where risk-averse subjects should bid more aggressively).

\begin{figure}[tp]
    \centering
    \includegraphics[scale=0.3]{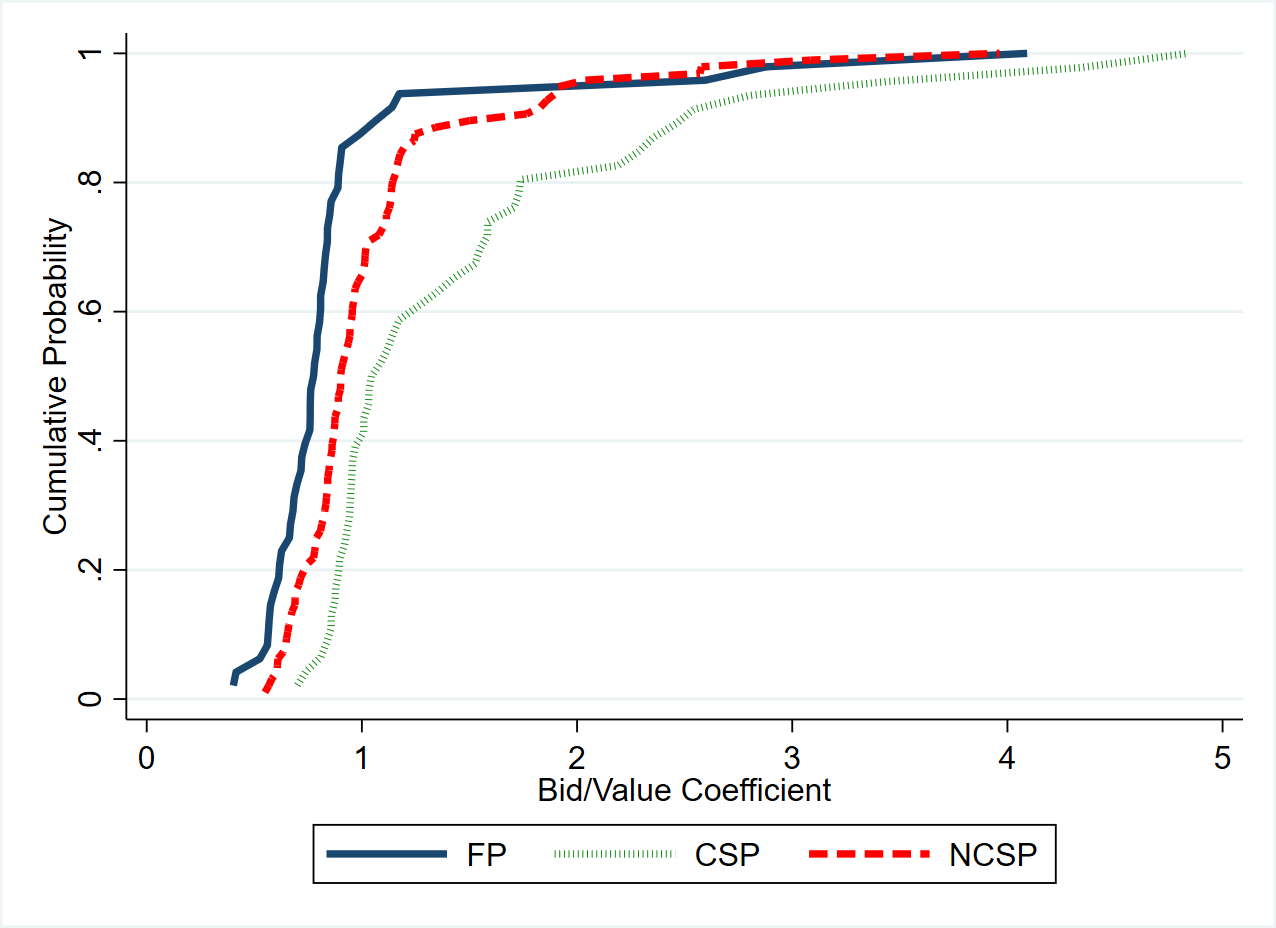}
    \caption{CDFs of bid/value coefficients}
    \label{fig:bidder_CDF}
\end{figure}

Finally, we analyze bidding behavior at the subject level.  The data for each bidder consists of 10 pairs of bids and values (one for each round of the experiment).  For each bidder, we calculate her \textit{bid/value coefficient} as the average of her 10 bid/value ratios.  The cumulative distribution functions (CDFs) of the bid/value coefficients are shown in Figure~\ref{fig:bidder_CDF}.  Several patterns are clear by inspection.  First, the fact that bid/value coefficients exceed 1 for many subjects reveals the prevalence of dominated strategies in the data.  Second, we observe a first-order stochastic dominance relationship when comparing the CDFs of either the FP or NCSP treatments to the CSP treatment.  Using a Kolmogorov-Smirnov test, we can also reject the null hypothesis that bid/value coefficients are drawn from the same underlying distribution in all three pairwise comparisons ($p < 0.01$ in all cases).  This further underscores the point that bidding behavior is significantly different across the three auction institutions.

\subsection*{Seller Behavior}

We now investigate the behavior of sellers.  In the credible auctions (FP and CSP treatments), the seller's role is passive.  Thus, our data consists only of the 48 sellers who participate in the NCSP treatment.  Recall that each seller makes two choices in each round: she decides which bidder wins the item and the price of the item.  We find that the seller selects the highest bidder as the winner in 99\% (476/480) of cases.  This behavior is consistent with both the dominant strategy of the seller and the rules of the auction.  Furthermore, the fact that sellers nearly always allocate the item to the highest bidder provides a basic rationality check and rules out subject confusion as an explanation for our results.

We now explore the degree to which sellers break the rules of the auction when making pricing decisions.  To do so, we define the "overcharging ratio" $s_{it}$ for seller $i$ in round $t$ of the experiment as

\begin{equation*}
    s_{it} = \frac{price_{it} - b^{min}_{it}}{b^{max}_{it} - b^{min}_{it}},
\end{equation*}

where $price_{it}$ is seller $i$'s chosen price in round $t$, $b^{max}_{it}$ is the highest bid in seller $i$'s group in round $t$, and $b^{min}_{it}$ is the lowest bid in seller $i$'s group in round $t$.\footnote{The overcharging ratio is undefined when $b^{max}_{it} = b^{min}_{it}$.  This only occurs in 10 out of the 480 auction markets in the NCSP treatment.  These 10 observations are omitted from the regressions shown in Table~\ref{table:seller_ratio}.}  The ratio $s_{it}$ captures the share of surplus that is extracted by seller $i$ in round $t$.  If $s_{it} = 0$, then the seller chooses a price equal to the lowest bid (i.e., she extracts no additional surplus beyond the amount prescribed by the rules of the auction).  If $s_{it} = 1$, then the seller chooses a price equal to the highest bid (i.e., she extracts the maximum possible surplus).\footnote{In our experiment, it is not possible for $s_{it} > 1$.  However, it is possible for $s_{it} < 0$ if the seller breaks the rules of the auction by undercharging.  This only happens in 2\% (9/470) of cases.}

We first investigate whether sellers' overcharging behavior changes with experience, and whether overcharging is correlated with a traditional measure of lying aversion taken from the literature.\footnote{To measure lying aversion, we use a task from \citet{gneezy}.  In particular, subjects observe a randomly generated integer from 1 to 10 (inclusive).  On a subsequent screen, they are asked to report the number that they saw on the previous screen and they are told that they will be paid exactly half the number that they report (in pounds).  Although the profit-maximizing choice is to report the maximum number of 10, we find that 90\% (43/48) of sellers in the NCSP treatment truthfully report their numbers.  We classify these 43 sellers as lie-averse.}  The results are shown in Regressions 1 - 2 of Table~\ref{table:seller_ratio}.  We find that sellers overcharge 29 percentage points more in the second half of the experiment than in the first half, and that lie-averse sellers overcharge 33 percentage points less than sellers who are not lie-averse.  Although both of these results are consistent with our intuition, only the latter effect is statistically significant at conventional levels.

\begin{table}[tp]
\centering
\begin{tabular}{lccc} \hline
& (1) & (2) & (3) \\ 
& Overcharging Ratio & Overcharging Ratio & Decision to Overcharge \\ \hline
$\mathbbm{1}_{\lbrace{Round > 5}\rbrace}$ & 0.289 &  \\
& (0.210) & & \\
$\mathbbm{1}_{\lbrace{Lie-Averse}\rbrace}$ & & -0.333** \\
& & (0.135) & \\
Bid Spread & & & 0.004*** \\
& & & (0.001) \\
Constant & 0.456** & 0.901*** & 0.690*** \\
& (0.218) & (0.039) & (0.058) \\
& & & \\
Observations & 470 & 470 & 470 \\ 
Number of subjects & 48 & 48 & 48 \\ \hline
\end{tabular}
\caption{OLS regressions of seller behavior in the NCSP auction \\ Standard errors are clustered at the subject level \\ Significance levels are indicated as follows: ** $p < 0.05$, *** $p < 0.01$}
\label{table:seller_ratio}
\end{table}

\begin{table}[tp]
    \centering
    \begin{tabular}{cccc}
    \toprule
        Seller Type & \# of Subjects & Average $s_i$ & Average $s_i \vert s_{it}>0$   \\
         \midrule
         Never Overcharge & 3 & 0 & - \\
         Sometimes Overcharge & 22 & 0.36 & 0.77 \\
         Always Overcharge & 23 & 0.91 & 0.91 \\
         \midrule
         All & 48 & 0.60 & 0.84 \\
         \bottomrule
    \end{tabular}
    \caption{Distribution of seller types in the NCSP auction}
    \label{table:seller_types}
\end{table}

We now investigate the prevalence of different seller types in our experiment.  In particular, we calculate the "overcharging coefficient" $s_i$ of seller $i$ as the average of her 10 overcharging ratios (one for each round of the experiment).  This allows us to classify sellers as one of three possible types: sellers who never overcharge, sellers who sometimes overcharge, and sellers who always overcharge.  Table~\ref{table:seller_types} shows the distribution of seller types in our experiment.  It is clear that overcharging is the norm: only three sellers never overcharge in all rounds of the experiment (i.e., $s_i = 0$).  We find that the 22 sellers who sometimes overcharge extract an average of 36\% of the available surplus, while the 23 sellers who always overcharge extract an average of 91\% of the available surplus.  This difference is both economically and statistically significant (Mann-Whitney test, $p < 0.001$).  Finally, of the 23 sellers who always overcharge, only six sellers consistently extract the maximum surplus (i.e., $s_i = 1$).

Taken together, our results suggest that most sellers in the NCSP auction are not profit-maximizing.  We now analyze whether there are any systematic patterns in sellers' deviations from profit-maximizing behavior.  We focus on two main questions.  First, do sellers who sometimes overcharge behave similarly to sellers who always overcharge, \textit{conditional on the decision to break the rules of the auction}?  We find that this is not the case.  Conditional on overcharging, sellers who sometimes overcharge extract an average of 77\% of the available surplus, which is still significantly different than the rate among sellers who always overcharge (Mann-Whitney test, $p = 0.010$).  Second, are sellers responsive to the size of the pecuniary gains associated with breaking the rules of the auction?  In principle, the seller's likelihood of overcharging the winner might be increasing in the difference between the maximum and minimum bids (i.e., the bid spread).  To test this hypothesis, we estimate an OLS regression of a dummy variable for overcharging (i.e., $s_{it} > 0$) on the bid spread.\footnote{The results are robust to using a probit regression.}  The results are shown in Regression 3 of Table~\ref{table:seller_ratio}.  We find a statistically significant effect: in particular, every 10 token increase in the bid spread increases the seller's probability of overcharging the winner by four percentage points.

\subsection*{Aggregate Outcomes}

We now compare efficiency and revenue across the different auction formats.  We use standard definitions: the outcome of an auction is efficient if the bidder with the highest value wins the item, and the revenue from an auction is the price of the item.  If subjects use symmetric and strictly increasing bidding functions, then all three auctions are efficient.  If these bidding functions are also an equilibrium, then all three auctions yield the same expected revenue for the seller.\footnote{To prove that the revenue equivalence theorem still holds in the NCSP auction, we only need to verify that the bidder with the highest valuation receives the object in equilibrium.  Recall that we restrict attention to the symmetric Perfect Bayesian equilibrium in undominated strategies.  Since bidders use symmetric and increasing bidding functions in equilibrium and the seller's only undominated strategy involves allocating the object to the highest bidder, this implies that the bidder with the highest valuation receives the object.}  In the case of two bidders with values independently and uniformly distributed on $[0,100]$, the expected revenue is 33.

\begin{table}[tp]
    \centering
    \begin{tabular}{cccccccccccc}
    \toprule
    & \multicolumn{5}{c}{Rounds 1 - 10} & & \multicolumn{5}{c}{Rounds 6 - 10} \\
    & & & & & & & & & & & \\
    & \multicolumn{5}{c}{Treatment} & & \multicolumn{5}{c}{Treatment}  \\
    Property & CSP & & FP & & NCSP & & CSP & & FP & & NCSP \\
    \midrule
    Efficiency & 0.80 & $<$ & 0.82 & $<$ & 0.84 & & 0.85 & $>^{**}$ & 0.79 & $<$ & 0.84 \\
    Theor. Efficiency & 1 & $=$ & 1 & $=$ & 1 & & 1 & $=$ & 1 & $=$ & 1 \\
    & & & & & & & & & & & \\
    Revenue & 36 & $<^{***}$ & 50 & $<^*$ & 53 & & 38 & $<^{**}$ & 49 & $<$ & 54 \\
    Theor. Revenue & 33 & $=$ & 33 & $=$ & 33 & & 33 & $=$ & 33 & $=$ & 33 \\
    \bottomrule
    \end{tabular}
    \caption{Average efficiency and revenue \\ Significance levels are indicated as follows: * $p < 0.10$, ** $p < 0.05$, *** $p < 0.01$}
    \label{table:properties}
\end{table}

Table~\ref{table:properties} documents the average efficiency and revenue in each treatment.\footnote{For the efficiency comparisons, we assess statistical significance using probit regressions of the variable of interest (i.e., a dummy variable for whether the bidder with the highest value wins the item) on treatment dummy variables.  For the revenue comparisons, we assess statistical significance using OLS regressions of the variable of interest (i.e., the price of the item) on treatment dummy variables.  In all cases, standard errors are clustered at the session level.}  We observe high rates of efficiency in our experimental auction markets: the fraction of efficient outcomes is at least 80\% in all three treatments.\footnote{If the object is allocated randomly, then 50\% of outcomes are efficient.  In all three treatments, we find that the fraction of efficient outcomes is significantly higher than random chance would predict (two-sided t-test, $p < 0.001$).}  Moreover, the differences in efficiency across the three treatments are not statistically significant.  This finding is also robust to subject experience.  When restricting attention to the last five rounds of the experiment, we observe similarly high rates of efficiency and only one significant treatment effect among the three pairwise comparisons.

We do observe significant differences in revenue across the three treatments.  In particular, we find that the CSP auction generates the least revenue while the NCSP auction generates the most revenue.  However, the only robust finding is the revenue inferiority of the CSP auction.  The additional revenue generated by the NCSP auction over the FP auction is not economically significant and also fails to be statistically significant in the last five rounds of the experiment.

\section{Underlying Mechanism}

Our experimental results show that the NCSP auction fails to converge to the FP auction.  Most sellers in the NCSP auction do not maximally overcharge the winning bidder.  Consequently, most bidders in the NCSP auction do not behave as though they are participating in a FP auction. 

In this section, we investigate the reason for this failure of theoretical equivalence.  First, we propose an alternative specification for the seller's preferences that includes an \emph{aversion to rule-breaking}.  We then show that this behavioral model generates predictions that are consistent with our experimental results.  Finally, we conduct an additional experimental treatment to test the underlying mechanism of the behavioral model.

\subsection{Behavioral Model}
We consider a behavioral seller who is \emph{averse to rule-breaking}.  Since we restrict attention to equilibria in undominated strategies, the seller would never choose a price below the second-highest bid.  Thus, we can characterize the price $t_y$ by a parameter $\alpha\in [0,1]$ as follows:
$$
t_y  = \alpha b_{(1)} + (1-\alpha) b_{(2)},
$$
where $b_{(1)}$ is the highest bid and $b_{(2)}$ is the second-highest bid.

For the seller, we allow for both a fixed cost and a variable cost of rule-breaking.  The fixed cost of rule-breaking is represented by the parameter $\gamma>0$.  The variable cost of rule-breaking is represented by the function
$$
c(\alpha)(b_{(1)} - b_{(2)}),
$$
where $c(\alpha)$ is a continuous, monotone, and concave function such that $c(0)=0$, $c'(0) \le 1$, and $c'(1) \ge 1$.  Then, the seller's utility function is as follows:
$$
u(\alpha) = \alpha b_{(1)} + (1-\alpha) b_{(2)} - c(\alpha)(b_{(1)} - b_{(2)}) - \gamma  \mathbbm{1}_{\lbrace{\alpha>0\rbrace}}.
$$

\noindent We can now update the theoretical predictions for the case of a behavioral seller.  Recall that $\mathbf{b^{FP}}$, $\mathbf{b^{NCSP}}$, $\mathbf{b^{CSP}}$ denote the symmetric and strictly increasing equilibrium bidding functions in the first-price, non-credible second-price, and credible second-price auctions, respectively.

\begin{prop}
\label{prop:BehavioralEquilibrium}
In equilibrium, 
\begin{itemize}
    \item [(i)] the seller selects the highest bidder as the winner.
    \item [(ii)] there exists $\hat\alpha \in [0,1]$ such that the seller's optimal strategy is
\begin{equation*}
\alpha^* =
\begin{cases}
    0 & \text{ if } b_{(1)} - b_{(2)} < \frac{\gamma}{\hat \alpha - c(\hat \alpha)} \\
    \hat \alpha & \text{ otherwise }
\end{cases},
\end{equation*}
where $\frac{\gamma}{\hat \alpha - c(\hat \alpha)}\ge 0$.
    \item [(iii)] for each bidder $i \in N$ and each value $\theta_i \in \Theta_i$, we have that \\
\begin{equation*}
\mathbb{E} \Big[\max_{j \neq i} \theta_j  \Big\vert \max_{j \neq i} \theta_j < \theta_i \Big] = \mathbf{b^{FP}}(\theta_i) \leq \mathbf{b^{NCSP}}(\theta_i) \le \mathbf{b^{CSP}}(\theta_i) = \theta_i.
\end{equation*}
\end{itemize}
\end{prop}

\begin{proof}
See Appendix \ref{sec:Proofs}.
\end{proof}

\noindent Proposition \ref{prop:BehavioralEquilibrium} provides two novel testable predictions.  First, it predicts that the equilibrium bidding function in the NCSP auction is nested between the equilibrium bidding functions of the FP and CSP auctions.  Intuitively, the NCSP auction is a convex combination of the other two auction formats, where the precise mixture depends on the seller's degree of rule-breaking aversion.  Second, it predicts that the seller will follow the rules of the auction when there are limited pecuniary gains from rule-breaking.

%Proposition \ref{prop:BehavioralEquilibrium} provides empirical implications for the NCSP and behavioral auctioneer. First of all, bidding in NCSP under the rule-breaking averse auctioneer is consistent with the observed bidding pattern. That is, the bid in the NCSP should lie between the bids for CSP and FP.

%On the seller's side, Proposition \ref{prop:BehavioralEquilibrium} also provides important empirical implications. The presence of variable costs of rule-breaking implies that the seller would not extract all surplus. That is, would charge the winner the price between the first- and second-highest bids as shown in Table \ref{table:seller_types}. The presence of the fixed costs of rule-breaking implies that the sellers would not always charge the price above as shown in Table \ref{table:seller_types}. Finally, Proposition \ref{prop:BehavioralEquilibrium} also generates the new testable implication. That is, the probability that the seller charges a price strictly above the second-highest bid should be correlated with the surplus to be extracted. 

\subsection{Revisiting the NCSP Auction}

In this section, we revisit the NCSP auction to demonstrate that the main predictions of the behavioral model are consistent with our experimental results. \\

\noindent \textbf{Prediction 1:} $\mathbf{b^{FP}}(\theta_i) \leq \mathbf{b^{NCSP}}(\theta_i) \le \mathbf{b^{CSP}}(\theta_i)$. \\

This pattern is reflected in our data.  Recall from Section 4 that the empirical bidding functions exhibit the predicted ordering and the estimated slopes are significantly different at the 1\% level.  We also find that 73\% (702/960) of bid-value pairs in the NCSP auction lie between the theoretical bounds of the FP and CSP auctions. \\

\noindent \textbf{Prediction 2:} For a sufficiently small bid spread $b_{(1)} - b_{(2)}$, $\alpha^* = 0$.  In other words, the seller chooses the price $t_y = b_{(2)}$. \\

In the NCSP auction, sellers choose a price equal to the second-highest bid in 20\% (96/480) of cases.  Furthermore, we find evidence that sellers are responsive to the bid spread when making their pricing decisions.  In particular, Regression 3 of Table~\ref{table:seller_ratio} shows that sellers are less likely to overcharge the winner when the bid spread is smaller, which is consistent with the qualitative predictions of the model. \\

%\noindent \textbf{Prediction 3:} The seller's chosen price $t_y$ is a function of both $b_{(1)}$ and $b_{(2)}$. \\

%Prediction 3 rules out simple heuristics where the seller chooses a price based on only the highest or the second-highest bids.  To test this, we run an OLS regression of the seller's chosen price on the highest and second-highest bids, with standard errors clustered at the subject level.  In line with the prediction, we find that the estimated coefficients on $b_{(1)}$ and $b_{(2)}$ are both positive and statistically significant ($p < 0.001$ and $p = 0.044$, respectively).

\subsection{Experimental Design}

While the predictions of the behavioral model are consistent with our experimental results, there are other possible forces at play.  For example, if the seller has other-regarding preferences, then she would also choose an intermediate price between the highest and second-highest bids.  In this section, we design a new experimental treatment to directly test the mechanism proposed in the behavioral model.  In particular, we define a "no-rules" auction (NR) that is procedurally identical to the NCSP auction but that differs in framing.  In both auctions, the seller has agency to determine the outcome of the auction (i.e., the winner and the price of the item).  However, the experimental instructions vary across the two treatments.  In the NCSP auction, both the rules of the auction and the seller's strategy space are described to subjects.  In the NR auction, the rules are omitted and only the seller's strategy space is described to subjects.\footnote{The instructions for the NR treatment are also included in the Online Appendix.}  Comparing sellers' pricing behavior across the NR and NCSP auctions allows us to pin down whether sellers' aversion to rule-breaking is the driving force behind our experimental results.

\subsubsection*{Implementation}

The additional treatment was conducted at the Laboratory for Economic and Decision Research (LEDR) at the University of East Anglia.  We recruited a total of 72 subjects across four sessions (with 18 subjects per session).  Each session lasted approximately 60 minutes.  The experiment was programmed and conducted with the z-Tree software \citep{fischbacher}.  Table~\ref{table:summarystatistics} provides more detailed summary statistics. 

\subsection{Experimental Results}

We first conduct a linear estimation of the bidding function in the NR auction (forcing the intercept to pass through zero).  Ordering the estimated bidding functions, we find that

\begin{equation*}
    \stackbelow{\hat{b}^{FP}(\theta_i)}{0.74\theta_i} <
    \stackbelow{\hat{b}^{NR}(\theta_i)}{0.75\theta_i} <^{***} 
    \stackbelow{\hat{b}^{NCSP}(\theta_i)}{0.83\theta_i}. 
\end{equation*}

\noindent Several insights emerge from these comparisons.  First, bidding behavior is significantly different between the NCSP and NR auctions.  This suggests that bidders believe that sellers will behave differently in the absence of auction rules.  Second, there is no significant difference in bidding behavior between the FP and NR auctions.  This allows us to draw a further inference: in the absence of auction rules, bidders' behavior is consistent with the belief that sellers will extract the maximum possible surplus.

We now compare the behavior of sellers between the NCSP and NR auctions.  To do so, we estimate OLS regressions of both the overcharging ratio ($s_{it}$) and a dummy variable for overcharging ($s_{it} > 0$) on a treatment dummy variable for the NR auction.  The results are shown in Table \ref{table:seller_ratio_2}.  We observe statistically significant differences in seller behavior on both the intensive and extensive margins.  In particular, sellers in the NR auction extract 27 percentage points more of the available surplus and are 14 percentage points more likely to overcharge the winner than in the NCSP auction.

Clearly, from the perspectives of both sides of the market, the NCSP and NR auctions are distinct mechanisms.  In the absence of auction rules, sellers' pricing behavior is closer to the FP auction.  Bidders, in turn, correctly anticipate this and behave as if they are participating in a FP auction.  Overall, the NR treatment demonstrates that sellers' aversion to rule-breaking plays a large role in explaining the results of the NCSP auction.

\begin{table}[tp]
\centering
\begin{tabular}{lcc} \hline
& (1) & (2) \\ 
& Overcharging Ratio & Decision to Overcharge \\ \hline
$\mathbbm{1}_{\lbrace{NR}\rbrace}$ & 0.269** & 0.137*** \\
& (0.122) & (0.044) \\
Constant & 0.603** & 0.817*** \\
& (0.116) & (0.041) \\
& & \\
Observations & 707 & 707 \\ 
Number of subjects & 72 & 72  \\ \hline
\end{tabular}
\caption{OLS regressions of seller behavior in the NCSP and NR auctions \\ Standard errors are clustered at the subject level \\ Significance levels are indicated as follows: ** $p < 0.05$, *** $p < 0.01$}
\label{table:seller_ratio_2}
\end{table}

\section{Conclusion}

We report results from a series of laboratory experiments that compare behavior and outcomes across different auction formats that vary in their credibility.  Our first main result is that, contrary to theory, the NCSP auction fails to converge to the FP auction.  While sellers often break the rules of the NCSP auction, they typically do not maximize revenue.  Consequently, bidders in the NCSP auction do not behave as though they are participating in a FP auction.  We propose an alternative specification for sellers' preferences that includes an aversion to rule-breaking.  We show that this behavioral model generates predictions that are consistent with our experimental results.  We then conduct an additional experimental treatment to test the underlying mechanism of the behavioral model.  In support of the model, we find that aversion to rule-breaking is able to organize the experimental data.

Our results have important implications for market design.  In particular, we have documented substantial heterogeneity in seller behavior in the NCSP auction.  This novel dimension of strategic uncertainty increases the burden of participation for bidders, who must now form beliefs about the seller's behavior in addition to considering the actions of other bidders.  This is a powerful argument for the use of credible auctions in the field, since they may be widely perceived as safer or simpler for bidders than non-credible auctions.\footnote{\citet{roth2008} cites "safety" as one of the essential components of a successful marketplace.}  In terms of practical advice, reforms that increase the transparency of auction institutions - such as requiring sellers to publicly disclose the highest and second-highest bids - could be helpful in mitigating bidders' concerns.  Indeed, \citet{hakimov} argue for the publication of priority score cutoffs to improve the verifiability of school choice mechanisms.  They also provide empirical support for the effectiveness of their suggestions in a laboratory experiment.

Finally, our experiment abstracts from participation decisions by randomly assigning each subject to a single auction format.  In real-life settings, bidders are likely to face a choice over auction formats and may exhibit a preference for credible auctions.  We believe that studying credible and non-credible auctions with endogenous entry is a promising area for future experimental research.

\newpage
\bibstyle{plainnat}
\bibliography{ref}

\clearpage

\appendix
\renewcommand{\theequation}{\Alph{section}.\arabic{equation}}
\setcounter{equation}{0}

\section{Proofs}
\label{sec:Proofs}

\begin{proof}[Proof of Proposition \ref{prop:BehavioralEquilibrium}]
{\bf (i)} This part follows trivially from the fact that sellers do not use dominated strategies.

\bigskip

\noindent
{\bf (ii)}
Taking the first-order conditions for the seller, we have that
$$
c'(\alpha) = 1.
$$
Given that $c''(\alpha) > 0$, $c'(0) \le 1$, and $c'(1) \ge 1$, there exists 
$$
\hat \alpha \in [0,1] \text{ such that } c'(\hat \alpha) = 1.
$$
However, the seller would only choose $\hat\alpha > 0$ if the utility under this decision is at least as large as the utility under $\alpha=0$.  That is,
$$
\alpha b_{(1)} + (1-\alpha) b_{(2)} - c(\alpha)(b_{(1)} - b_{(2)}) - \gamma  \mathbbm{1}_{\alpha>0} \ge b_{(2)}.
$$
Simplifying the above expression, we can see that the inequality is satisfied if and only if
$$
b_{(1)} - b_{(2)} \ge \frac{\gamma}{\hat \alpha - c(\hat\alpha)}.
$$
Given our assumptions on $c'(\alpha)$ and the fact that $c(0) = 0$, this guarantees that $\hat \alpha - c(\hat\alpha) > 0$.

\bigskip

\noindent
{\bf (iii)}
For this purpose, we need to calculate the expected price in the NCSP auction.  The expected price can be expressed as follows:
\begin{align*}
\mathbb{E}(t_y)  & = \hat\alpha F^{n-1}(\bar\theta) \hat b(\theta) + \int_{0}^{\theta} b(\xi) d F^{n-1}(\xi) - \alpha \int_{0}^{\bar\theta} b(\xi) dF^{n-1}(\xi) =
\\ & =
\alpha \int_{0}^{\bar\theta} \left(b(\theta) - b(\xi) \right) dF^{n-1}(\xi) + \int_0^{\theta} b(\xi) dF^{n-1}(\xi),
\end{align*}
where
$$
\bar\theta = b^{-1} \left(b_{(1)} - \frac{\gamma}{\hat\alpha - c(\hat\alpha)} \right)
$$
is the threshold value at which the seller stops overcharging the winner.

\medskip

\noindent
{\bf ($\mathbf{b^{FP}}(\theta_i) \leq \mathbf{b^{NCSP}}(\theta_i)$)}
Since $b(\xi)\le b(\theta)$, we have that
\begin{align*}
    \alpha \int_{0}^{\bar\theta} \left(b(\theta) - b(\xi) \right) dF^{n-1}(\xi) & + \int_0^{\theta} b(\xi) dF^{n-1}(\xi) \le \int_{0}^{\theta}b(\xi)dF^{n-1}(\xi) 
    \le \\ & \le   \int_{0}^{\theta}b(\theta)dF^{n-1}(\xi) = F^{n-1}(\theta)b(\theta).
\end{align*}

\noindent At the same time, the revenue equivalence theorem implies that 
$$
\alpha \int_{0}^{\bar\theta} \left(b(\theta) - b(\xi) \right) dF^{n-1}(\xi) + \int_0^{\theta} b(\xi) dF^{n-1}(\xi) = F^{n-1}(\theta)\mathbf{b^{FP}}(\theta).
$$
Therefore,
$$
F^{n-1}(\theta)\mathbf{b^{FP}}(\theta) \le F^{n-1}(\theta)b(\theta),
$$
which yields
$$
\mathbf{b^{FP}}(\theta) \le b(\theta).
$$

\medskip

\noindent
{\bf ($\mathbf{b^{NCSP}}(\theta_i) \leq \mathbf{b^{CSP}}(\theta_i)$)}
The revenue equivalence theorem implies that $\mathbb{E}(t_y) = \mathbb{E}(b^{CSP})$.  That is,
\begin{align*}
\alpha \int_{0}^{\bar\theta} \left(b(\theta) - b(\xi) \right) dF^{n-1}(\xi) + \int_0^{\theta} b(\xi) dF^{n-1}(\xi) =  \int_0^{\theta} \xi dF^{n-1}(\xi).
\end{align*}
Since $\bar \theta \le \theta$ and $b(\theta)$ is an increasing function, we have that
$$
\alpha \int_{0}^{\bar\theta} \left(b(\theta) - b(\xi) \right) dF^{n-1}(\xi) \ge 0.
$$
Therefore,
$$
\int_0^{\theta} b(\xi) dF^{n-1}(\xi) \le  \int_0^{\theta} \xi dF^{n-1}(\xi).
$$
Finally, if $\gamma$ is sufficiently low, then $\bar\theta>0$ and we are able to conclude that 
$$
\int_0^{\theta} b(\xi) dF^{n-1}(\xi) <  \int_0^{\theta} \xi dF^{n-1}(\xi).
$$

\end{proof}
%\clearpage
%\section{Experimental Instructions (Online Only)}
%\label{sec:AppendixInstructions}
%\includepdf[pages=-]{experimental_instructions}

\end{document}